\documentclass{article}

\usepackage{microtype}
\usepackage{graphicx}
\usepackage{subfigure}
\usepackage{booktabs} 

\usepackage{hyperref}
\usepackage{tabularx}
\usepackage{colortbl} 
\usepackage{array}    
\usepackage{makecell}
\usepackage{multirow} 
\usepackage{multicol} 
\usepackage{makecell}

\usepackage[accepted]{icml2025}

\usepackage{amsmath}
\usepackage{amssymb}
\usepackage{mathtools}
\usepackage{amsthm}

\usepackage[capitalize,noabbrev]{cleveref}

\definecolor{headercolor}{rgb}{0.824, 0.804, 0.784}
\definecolor{headercolordark}{rgb}{0.86, 0.63, 0.51}
\definecolor{roweven}{rgb}{0.96, 0.94, 0.92}
\definecolor{rowodd}{rgb}{0.92, 0.90, 0.88}

\theoremstyle{plain}

\theoremstyle{definition}

\theoremstyle{remark}

\usepackage[textsize=tiny]{todonotes}

\icmltitlerunning{COMPASS - Comparison Of Models using Probabilistic Assessment in Simulation-based Settings}

\begin{document}

\twocolumn[
\icmltitle{A COMPASS to Model Comparison and Simulation-Based Inference\\in Galactic Chemical Evolution}




\begin{icmlauthorlist}
\icmlauthor{Berkay Günes}{yyy}
\icmlauthor{Sven Buder}{anu,arc}
\icmlauthor{Tobias Buck}{yyy,comp}
\end{icmlauthorlist}

\icmlaffiliation{yyy}{Interdisciplinary Center for Scientific Computing (IWR), University of Heidelberg,
 Im Neuenheimer Feld 205, D-69120 Heidelberg, Germany}
\icmlaffiliation{comp}{Universität Heidelberg, Zentrum für Astronomie, Institut für Theoretische Astrophysik, Albert-Ueberle-Straße 2, D-69120 Heidelberg, Germany}
\icmlaffiliation{anu}{Research School of Astronomy and Astrophysics, Australian National University, Canberra, ACT 2611, Australia}
\icmlaffiliation{arc}{ARC Centre of Excellence for All Sky Astrophysics in 3 Dimensions (ASTRO 3D), Australia}

\icmlcorrespondingauthor{Berkay Günes}{b.guenes@stud.uni-heidelberg.de}
\icmlcorrespondingauthor{Tobias Buck}{tobias.buck@iwr.uni-heidelberg.de}

\icmlkeywords{Model comparison, simulation-based inference, galactic chemical evolution}

\vskip 0.3in
]



\printAffiliationsAndNotice{}  

\begin{abstract}
We present \texttt{COMPASS}, a novel simulation-based inference framework that combines score-based diffusion models with transformer architectures to jointly perform parameter estimation and Bayesian model comparison across competing Galactic Chemical Evolution (GCE) models. \texttt{COMPASS} handles high-dimensional, incomplete, and variable-size stellar abundance datasets.
Applied to high-precision elemental abundance measurements, \texttt{COMPASS} evaluates 40 combinations of nucleosynthetic yield tables. The model strongly favours Asymptotic Giant Branch yields from NuGrid and core-collapse SN yields used in the IllustrisTNG simulation, achieving near-unity cumulative posterior probability. Using the preferred model, we infer a steep high-mass IMF slope and an elevated Supernova\,Ia normalization, consistent with prior solar neighbourhood studies but now derived from fully amortized Bayesian inference.
Our results demonstrate that modern SBI methods can robustly constrain uncertain physics in astrophysical simulators and enable principled model selection when analysing complex, simulation-based data.
\end{abstract}

\section{Motivation}
\label{motivation}
Cosmological simulations of galaxy formation \citep[e.g.][]{Sawala2016,Hopkins2018,Pillepich2018,Buck2020,Buck2020c,Font2020,Agertz2021} rely on a small set of galactic parameters to capture key processes in stellar evolution, such as star formation and supernova feedback. Two particularly uncertain inputs are the shape of the initial mass function (IMF), which determines the mass distribution of stars formed from the interstellar medium (ISM), and the rate and delay-time distribution of Type Ia supernovae (SN\,Ia).

These parameters strongly influence chemical enrichment histories \citep{2005A&A...430..491R,2015MNRAS.449.1327V,2015MNRAS.451.3693M}, yet remain poorly constrained by observations. Moreover, GCE models vary significantly in their choice of nucleosynthetic yields and assumptions about enrichment sources, including asymptotic giant branch (AGB) stars, core-collapse supernovae (cc-SNe), and SN\,Ia.
For instance, high-mass IMF slopes steeper than canonical values have been proposed by various studies \citep[Tab.~7]{2016ApJ...824...82C,2015ApJ...806..198W,Rybizki2015,2014ApJ...796...75C}. Likewise, SN\,Ia normalization and delay-time distribution remain active areas of debate \citep{2010ApJ...722.1879M,2012MNRAS.426.3282M,2015ApJ...810..137J,Buck2021}. Different stellar evolution models also yield a broad range of element production rates \citep[e.g.][]{1997NuPhA.621..467N,Kobayashi_2006,portinari,2010MNRAS.403.1413K,2014MNRAS.437..195D,2014ApJ...797...44F,2016ApJ...825...26K}.

Here, we introduce \texttt{COMPASS}, a scalable simulation-based inference (SBI; \citealp{Cranmer2020}) framework to overcome these challenges.
Building on the work of \citet{Philcox_2019}, who used computationally expensive Hamiltonian Monte Carlo (HMC) methods, \texttt{COMPASS} is designed for robust Bayesian model comparison and parameter inference from large, complex stellar abundance datasets.
\section{Related Work}
%
Recent work has increasingly applied machine learning to model comparison in astrophysics. \citet{karchev2023simsims} used a deep learning method for Bayesian model comparison \citep{elsemüller2023deep} to evaluate simulation-based SN\,Ia light curve models. \citet{zanisi2021deep} compared Illustris \citep{vogelsberger2014introducing} and IllustrisTNG \citep{Pillepich2018} simulations to $r$-band Sloan Digital Sky Survey (SDSS) images \citep{kollmeier2019sdss} by using two PixelCNNs \citep{van2016pixel} to generate pixel-wise anomaly scores.
\citet{jin2024quantitatively} applied GANomaly \citep{akcay2019ganomaly}, an anomaly detection model based on Generative Adversarial Networks (GANs; \citealp{goodfellow2020generative}), to evaluate NIHAO galaxy simulations \citep{wang2015nihao,buck2019nihao,Buck2020c} via anomaly scores relative to SDSS images. Similarly, \citet{Zhou2024} combined out-of-distribution detection with amortized Bayesian model comparison to assess simulated galaxy images against SDSS observations.
\section{Data}
\paragraph{Observational Data and Elemental Abundances}
The chemical composition of stars is typical measured in logarithmic abundance ratios of the number fraction of elements $[\mathrm{X}/\mathrm{Fe}]$ and $[\mathrm{Fe}/\mathrm{H}]$, defined as:
$ [\mathrm{X}/\mathrm{Y}] = \log_{10}(N_\mathrm{X}/N_\mathrm{Y})_\mathrm{star} - \log_{10}(N_\mathrm{X}/N_\mathrm{Y})_\odot,$ 
where $N_\mathrm{X}$ is the number density of element X, and $\odot$ refers to solar abundances from \citet{2009ARA&A..47..481A}.

Real data is taken from the high-precision stellar abundance measurements from \citet{Nissen_2020}, which provide detailed compositions of solar-type stars. This dataset offers a clean, well-characterized sample for evaluating \texttt{COMPASS} inference fidelity.

\paragraph{Galactic Chemical Evolution Model and Mock Data}
We use the \texttt{CHEMPYMulti} simulator \citep{Philcox_2019}, based on the original \texttt{CHEMPY} GCE model \citep{Rybizki_2017}, to generate stellar abundance predictions given parameterized stellar and ISM physics. The model evolves chemical abundances over cosmic time using published yield tables for SN\,Ia, SN\,II, and AGB feedback \citep{Philcox_2018}.
These nucleosynthetic yield tables are theoretical predictions that quantify the mass of each chemical element produced and ejected by a star of a given initial mass and metallicity over its lifetime.
Our inference operates over six free parameters, grouped as: (1) \textbf{Global parameters} ($\vec{\Lambda}$): the IMF slope $\alpha_\mathrm{IMF}$ \citep{2003PASP..115..763C} and the SN\,Ia normalization $\log_{10}(N_\mathrm{Ia})$, assumed constant across stars. 
(2) \textbf{Local parameters} ($\{\vec{\Theta}_i\}$): star-formation efficiency $\log_{10}(\mathrm{SFE})$, SFR peak time $\log_{10}(\mathrm{SFR}_\mathrm{peak})$, and outflow fraction $x_\mathrm{out}$, each specific to a star's birth environment \citep{Rybizki_2017, Philcox_2019}. 
(3) \textbf{Birth times} ($\{T_i\}$): formation times in Gyr, setting ISM conditions from which stellar abundances are drawn.
Because H is used for normalization, only $n_\mathrm{el}=8$ independent elemental abundances are required for modelling with \texttt{Chempy}.
To simulate observational uncertainties, we perturb synthetic abundances with $5\%$ Gaussian noise.

\paragraph{GCE Models for comparison}
We compare 40 different combinations of nucleosynthetic yield tables, spanning a range of theoretical prescriptions for AGB and core-collapse SN yields. These combinations span literature-derived SN\,Ia, SN\,II, and AGB yields \citep[e.g.][]{1997NuPhA.621..467N, Kobayashi_2006, portinari, 2010MNRAS.403.1413K, 2014MNRAS.437..195D, 2014ApJ...797...44F, Karakas2016}.
A full list of combinations is provided in Table~\ref{table:yield-comparison-results} in the Appendix. The goal is to identify which yield models most accurately reproduce the observed chemical abundance patterns, based on the posterior model probability $P(\mathcal{M}_k \mid \mathbf{x})$.
\section{Methods}
\begin{figure*}
    \centering
    \includegraphics[width=0.9\textwidth]{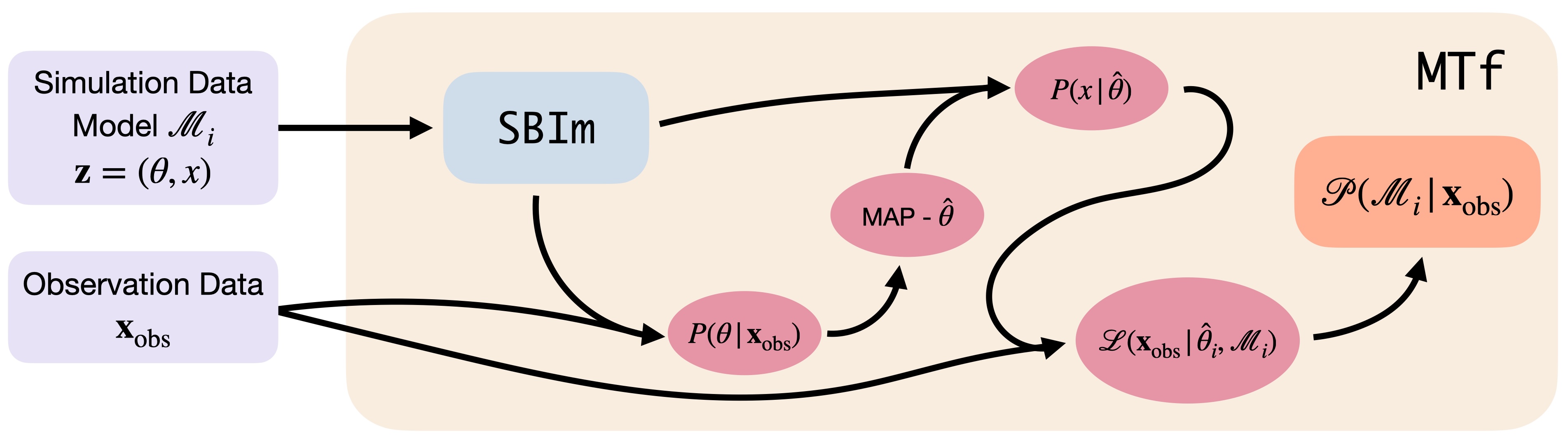}
\caption{Flow chart of our model comparison workflow. Each of the $40$ candidate models ($\mathcal{M}_i$) is used to train a separate Score-Based Inference Model (SBIm). For a given observation $\mathbf x_{\rm obs}$, the corresponding trained \texttt{SBIm} infers the posterior $P(\theta|\mathbf x_{\rm obs})$ to find the MAP parameters $\hat\theta$. The same model is then used to estimate the maximized likelihood $L(\mathbf x_{\rm obs}|\hat\theta_i,\mathcal M_i)$. The set of likelihoods from all $40$ models is then used to derive the final model posterior probabilities, allowing for a principled comparison of the underlying physical assumptions.}
\label{fig:flow-chart}
\end{figure*}
Our goal is to compare GCE models based on how well they reproduce observed stellar abundance data. Each competing model $\mathcal{M}_i$ corresponds to a different nucleosynthesis yield set implemented in the \textsc{Chempy} simulator. We frame this as a Bayesian model comparison problem and use SBI to estimate posterior model probabilities.
A key challenge in stellar data is the variable and potentially large number of observations. To address this, we design a machine learning pipeline (Figure \ref{fig:flow-chart})  capable of handling sets of varying cardinality by combining score-based diffusion models with a transformer architecture \citep[][]{peebles2023scalablediffusionmodelstransformers,gloeckler2024allinonesimulationbasedinference}.
\paragraph{Simulation-Based Inference}
SBI methods \citep[e.g.][]{Cranmer2020, Papamakarios:2021, gloeckler2024allinonesimulationbasedinference} estimate posteriors without requiring an explicit likelihood. Given a generative model $\mathcal{M}$ with parameters $\theta$ and synthetic observations $\Vec{X}$ (e.g., simulated stellar abundances), we train a neural network to approximate $p(\theta \mid \Vec{X}, \mathcal{M})$. Once trained, this posterior estimator can be applied to real observations $\Vec{X}_\mathrm{R}$.
Our implementation uses a conditional diffusion model trained on joint pairs $(\theta, \Vec{x})$, along with a binary condition mask $\mathcal{M}_C$ indicating which values are observed or latent. This allows us to operate as both a Neural Likelihood Estimator (NLE) and Neural Posterior Estimator (NPE). 
From the posterior samples, we estimate MAP parameters $\hat{\theta}$ and then generate samples from the likelihood $p(\Vec{x} \mid \hat{\theta})$ by inverting the mask $\mathcal{M}_C$. A Gaussian kernel density estimator (KDE) is used to approximate the likelihood at the observed $\Vec{x}$, enabling model comparison via KDE-estimated log-likelihoods.
\paragraph{Score-Based Transformer Architecture}
To model the score function, we follow \citet{gloeckler2024allinonesimulationbasedinference} and use a transformer-based diffusion model. Transformers overcome limitations of feed-forward networks in handling set-structured or sequential data. We use the adaLN-Zero DiT architecture as proposed in \citet{peebles2023scalablediffusionmodelstransformers}, adapted for continuous data.
An attention mask prevents latent tokens (drawn from noise) from attending to each other, focusing instead on observed inputs. Inputs are embedded in a high-dimensional space; diffusion timesteps are encoded via a sinusoidal timestep embedding \citep{vaswani2023attentionneed}. The condition mask $\mathcal{M}_C$ guides the network to infer unobserved values based on observed ones. More details are given in Appendix~\ref{sec:calibration}.
\paragraph{Training}
Each of the $40$ models is trained on $10^6$ simulated pairs and validated on $10^5$ examples. Parameters $(\Lambda, \Theta_i)$ are sampled from a uniform prior spanning a range of $\pm 5\sigma_{\text{Prior}}$ centred on the prior means $\mu_{\text{Prior}}$, following the setup of \citet{Rybizki_2017}. This wide support mitigates bias toward the prior.
To simulate realistic measurement errors, we perturb the synthetic abundances with $5\%$ Gaussian noise. This value is chosen to be representative of the typical $1-\sigma$ statistical uncertainties reported in the \citet[][Table 3]{Nissen_2020} dataset. Further details on hyperparameter tuning, diffusion scheduling, and calibration are provided in Appendix~\ref{sec:calibration}.
\paragraph{Model Comparison}
In simulation-based settings where the likelihood is implicit and model evidence is unavailable, we perform model comparison using the Akaike Information Criterion (AIC) \citep{AKAIKE19781877, burnham2002model}. 
AIC allows us to estimate the relative quality of models based on their maximized log-likelihood $\ln\mathcal{L}(\mathbf{x} \mid \hat{\theta}_i, \mathcal{M}_i)$ and is suitable for non-nested models derived from simulators.
To compare a set of candidate models $\{\mathcal{M}_i\}$, we first compute AIC differences $\Delta_i(\text{AIC}) = \text{AIC}_i - \text{AIC}_{\min}$, where $\text{AIC}_{\min}$ is the minimum AIC value among the models.
The relative likelihood of model $\mathcal{M}_i$ given the data $\mathbf{x}$ is then proportional to $\exp(-\frac{1}{2} \Delta_i(\text{AIC}))$.
Normalizing these relative likelihoods yields posterior model probabilities \citep{Wagenmakers2004, AKAIKE1979, Bozdogan_1987}.
As detailed in Appendix~\ref{sec:AIC}, when all compared models have the same number of parameters (as is the case for the yield set comparisons in \texttt{COMPASS}), these probabilities simplify to:
\begin{align}
    \mathcal{P}(\mathcal{M}_i \mid \mathbf{x}) = \frac{\mathcal{L}(\mathbf{x} \mid \hat{\theta}_i, \mathcal{M}_i)}{\sum_j \mathcal{L}(\mathbf{x} \mid \hat{\theta}_j, \mathcal{M}_j)}.
\end{align}
This softmax-based comparison ranks models by their plausibility given the data. 
To assess whether a preferred model $\mathcal{M}_j$ is statistically supported, we perform hypothesis testing using the Bayes factor relative to a null model $H_0$, following \citet{MOREY20166}. Details are provided in Appendix~\ref{sec:significance}.
\begin{figure}[t]
    \centering
    \includegraphics[width=1\linewidth]{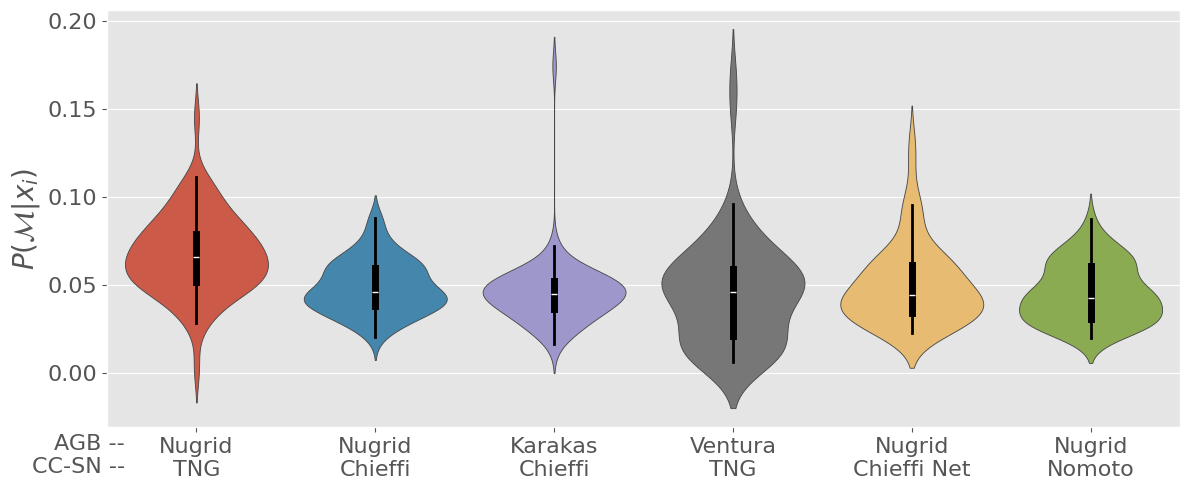}
    \includegraphics[width=1\linewidth]{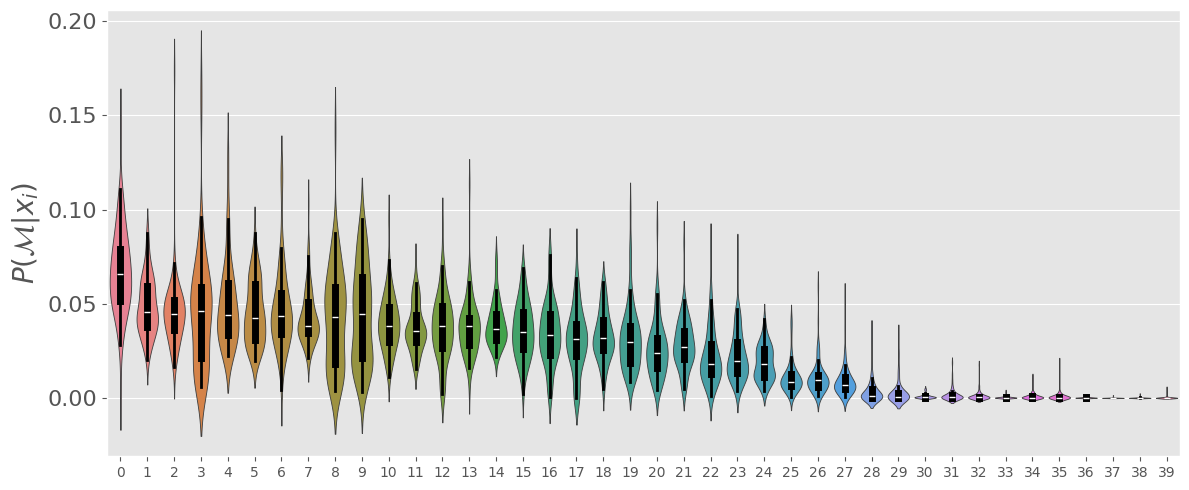}
    \includegraphics[width=1\linewidth]{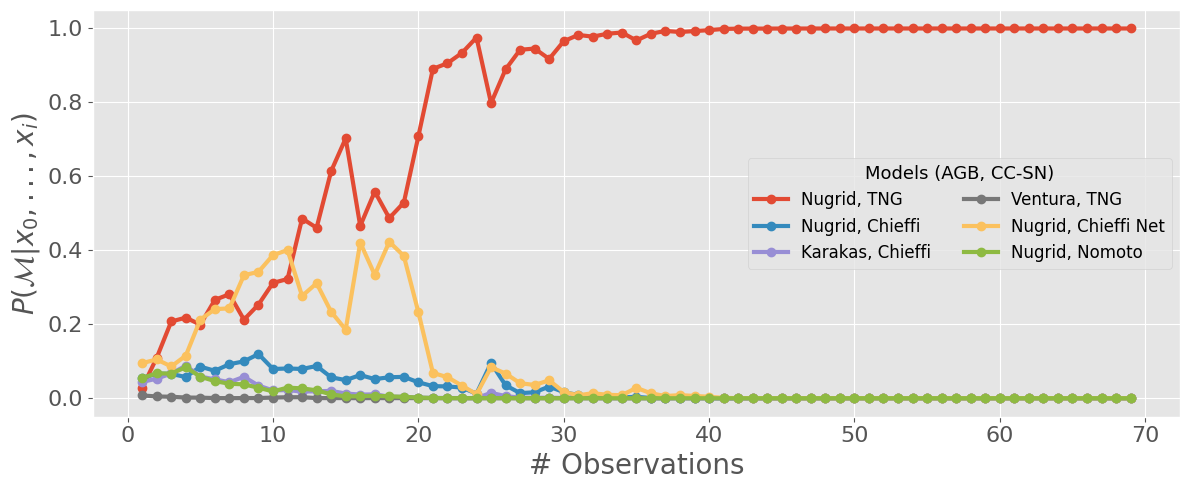}
\vspace{-2em}
\caption{Observational Yield-Set Inference
Bayesian comparison of $40$ nucleosynthetic yield set combinations using observational data from \citet{Nissen_2020}.
\textbf{Top:} Top six single-star model posteriors $P(\mathcal{M}_k|x_i)$. 
\textbf{Middle:} Single-star posterior probabilities for all $40$ tested combinations. 
\textbf{Bottom:} Cumulative model posterior probability $P(\mathcal{M}_k|x_0,\dots,x_i)$.
} \label{fig:MTf-nissen}
\end{figure}
\section{Results}
The \texttt{COMPASS} framework is applied to observational data to address two primary astrophysical objectives:
first, performing Bayesian model comparison across competing Galactic Chemical Evolution (GCE) models, and second, inferring key galactic parameters using the model(s) identified as most plausible. 
The rigorous evaluation of \texttt{COMPASS}'s performance, including its consistent recovery of ground-truth models and parameters on mock datasets, is presented in Appendix~\ref{section:Results.Yields.Chempy}.
\paragraph{Model Comparison}
We apply \texttt{COMPASS} to individual stellar abundances from the \citet{Nissen_2020} dataset, comparing 40 GCE models based on different combinations of AGB and CC-SN nucleosynthetic yield tables. These posterior model probabilities are then used to rank yield sets and guide downstream parameter inference.

Figure~\ref{fig:MTf-nissen} summarizes the model comparison results: Top panel: Violin plots of single-star model posteriors $P(\mathcal{M}_k \mid x_i)$ for selected competitive yield combinations show strong preference for specific models on a per-star basis. 
Middle panel: All 40 yield sets, sorted by median model posterior across stars, reveal overall model performance distribution.
Bottom panel: The cumulative posterior probability $P(\mathcal{M}_k \mid x_0, \dots, x_i)$ increases as more stars are considered, highlighting the growing statistical significance.

A clear preference emerges for models combining \citet{Ritter_2018} (NuGrid) AGB yields with the CC-SN yields used in the IllustrisTNG simulation \citep{Pillepich2018, Kobayashi_2006, portinari1997galacticchemicalenrichmentnew}. This combination achieves a relative cumulative posterior probability nearing $100\%$ after incorporating all 69 stars. While some models perform well for individual stars, their posterior probability drops when aggregated over the full dataset. A full ranking is provided in Table~\ref{table:yield-comparison-results} (Appendix).
These results underscore the strength of \texttt{COMPASS} in distinguishing physically motivated GCE models using real observational data, enabling robust model selection for chemical evolution studies.
\paragraph{Inferring Galactic Parameters} \label{sub-section:Results.Params}
Before applying \texttt{COMPASS} to real data, we benchmarked it against both the SBI pipeline from \citet{buck2025inferringgalacticparameterschemical} and Hamiltonian Monte Carlo (HMC) inference used in \citet{Philcox_2019}. Across tests involving matched and mismatched yield sets—including more complex data from the IllustrisTNG simulation—\texttt{COMPASS} reliably recovered ground-truth parameters (Appendix~\ref{sec:inference}).

\begin{figure}[!htb]
    \centering
    \includegraphics[width=1\linewidth]{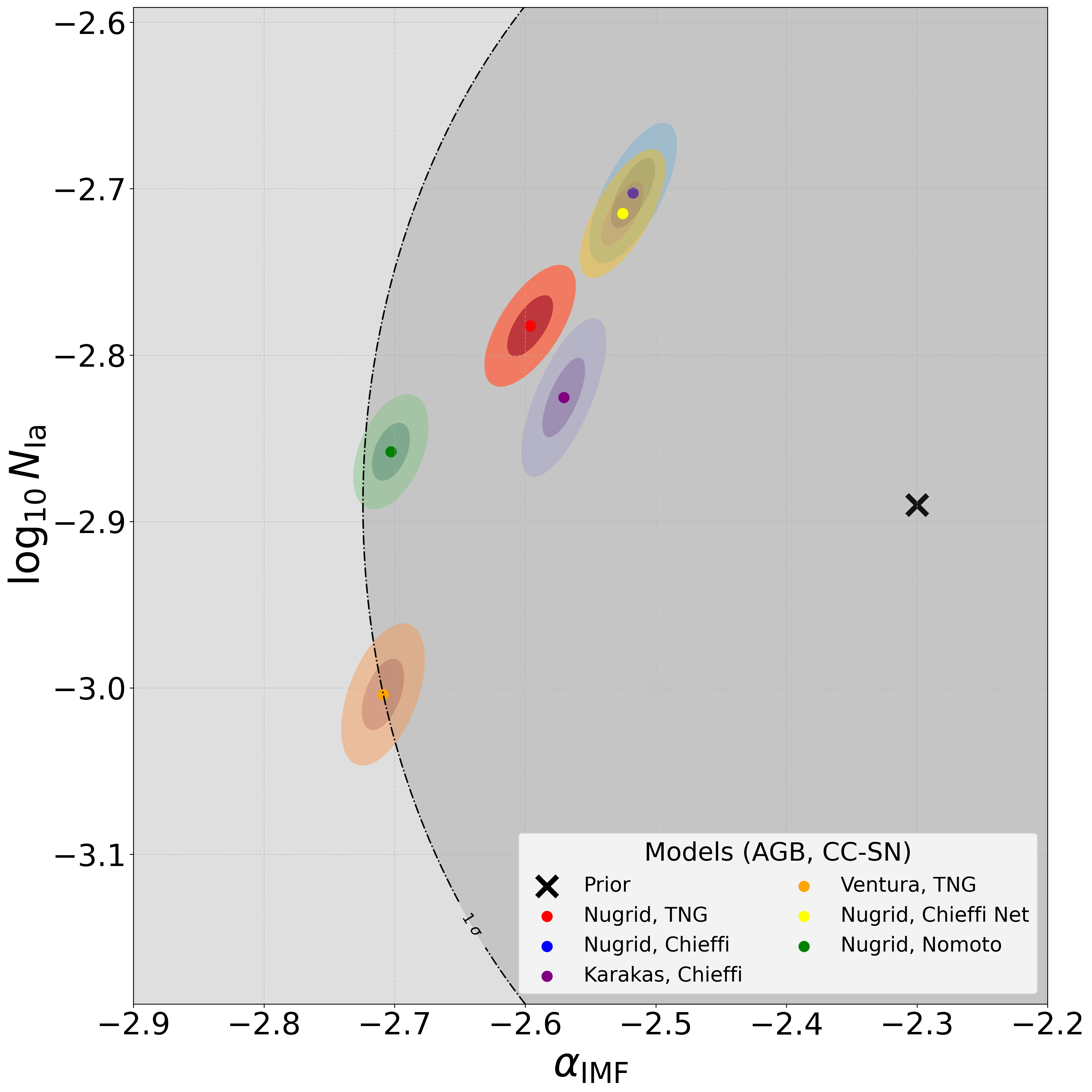}
    \vspace{-2em}
    \caption{Inferred Galactic Parameters from Observational Data
Joint posterior contours ($1\sigma, 2\sigma$) for the IMF high-mass slope $\alpha_{\rm{IMF}}$ and SN Ia normalization $\log_{10}N_{\rm{Ia}}$, from \citet{Nissen_2020} using \texttt{COMPASS}.
Colors denote the top six yield set combinations identified in Fig.~\ref{fig:MTf-nissen} and discussed in Tab.~\ref{table:nissen-inference-results}).
The black 'X' and dashed line indicate the \texttt{CHEMPY} prior mean and standard deviation, respectively.} 
\label{fig:Nissen-Inference}
\end{figure}

We then used the best-performing yield sets to infer global parameters from the \citet{Nissen_2020} sample. Figure~\ref{fig:Nissen-Inference} shows the joint posterior distributions of the IMF slope $\alpha_{\rm IMF}$ and the SN\,Ia normalization $\log_{10} N_{\rm Ia}$ for the top six yield combinations. Table~\ref{table:nissen-inference-results} in the Appendix summarizes their means and uncertainties.
Despite minor shifts depending on the yield set, the top-performing models yield consistent constraints with posterior distributions that are well-separated from prior means, indicating strong data-driven constraints. The best model (NuGrid AGB + TNG CC-SN yields) gives $\alpha_{\rm IMF} = -2.60 \pm 0.02$ and $\log_{10} N_{\rm Ia} = -2.78 \pm 0.02$. 
These values point to a systematically steeper IMF slope than the canonical Salpeter value of $-2.35$ or the Chabrier prior mean of $-2.3$, and a SN\,Ia normalization significantly elevated relative to the prior mean of $-2.89$ consistent with earlier studies of the solar neighborhood \citep[e.g.,][]{Rybizki2015}.
\paragraph{Scientific Impact}
Our results demonstrate that combining high-fidelity observational data with SBI and diffusion-based inference enables precise discrimination among theoretical models of stellar nucleosynthesis. \texttt{COMPASS} not only selects the most plausible yield prescriptions but also delivers tight, interpretable posteriors on key galactic parameters—$\alpha_{\rm IMF}$ and $N_{\rm Ia}$—critical for understanding galaxy-scale feedback and enrichment. This framework provides a powerful tool for constraining uncertain physics in cosmological simulations and stellar population synthesis.
\section{Conclusion and Summary}
We introduced \texttt{COMPASS}, a simulation-based inference framework that performs joint model comparison and parameter estimation. Here, we have employed it in the context of Galactic Chemical Evolution (GCE). Built on a score-based diffusion model and transformer architecture, \texttt{COMPASS} handles high-dimensional, partially observed data while maintaining scalability and statistical rigor.

Our method was first validated on synthetic data, where it reliably recovered ground-truth models and parameters. Applied to real stellar abundance data from \citet{Nissen_2020}, \texttt{COMPASS} identified a strongly preferred nucleosynthetic yield combination—NuGrid AGB yields and TNG CC-SN yields—and inferred global galactic parameters with high precision. The results favor an IMF slope significantly steeper than canonical models and an elevated SN\,Ia normalization, both of which have critical implications for galaxy formation and enrichment modeling.

\texttt{COMPASS} represents a novel integration of modern simulation-based inference techniques with physically motivated astrophysical models. Unlike traditional MCMC-based GCE studies, our framework enables efficient, amortized inference and principled Bayesian model comparison across a large space of competing simulators. 

Looking ahead, we anticipate that \texttt{COMPASS} can revolutionize the modeling of chemical evolution in galaxies. By enabling rigorous, data-driven selection among yield prescriptions and accurate inference of physical parameters, it offers a scalable pathway to calibrate and test subgrid physics in cosmological simulations. Future extensions to hierarchical inference and time-resolved abundance datasets could further improve our understanding of galactic feedback and chemical enrichment over cosmic time.
%
%
\section*{Impact Statement}
The authors are not aware of any immediate ethical or societal implications of this work. This work purely aims to aid scientific research and proposes a method for SBI in challenging astrophysical settings.
While there will certainly be many potential societal consequences of our work, none of which we feel must be specifically highlighted here.
%
%
\bibliography{references}
\bibliographystyle{icml2025}

\newpage
\appendix
\onecolumn

\section{Derivation of Model Posterior}
\label{sec:AIC}
Model comparison typically relies on model evidence. 
However, in simulator-based settings where the likelihood function is often intractable, calculating model evidence directly is not feasible.
An alternative approach for model comparison, particularly suitable for non-nested models, is to use the Akaike Information Criterion (AIC) \cite{AKAIKE19781877}. 
AIC estimates the prediction error and thereby the relative quality of statistical models for a given set of data.

Given the maximized log-likelihood for model $\mathcal{M}_i$, denoted as $\ln\mathcal{L}(\mathbf{x}|\hat{\theta}_i, \mathcal{M}_i)$ (where $\hat{\theta}_i$ are the parameter values that maximize the likelihood for model $\mathcal{M}_i$), the AIC is defined as:
\begin{align}
    \text{AIC}_i = -2\ln\mathcal{L}(\mathbf{x}|\hat \theta_i, \mathcal{M}_i) + 2k_i \label{eq:aic_def}
\end{align}
where $k_i$ is the number of estimable parameters in model $\mathcal{M}_i$. 
The model with the lowest AIC is generally preferred.

To compare a set of $N$ models, we first calculate the AIC difference for each model $\mathcal{M}_i$ relative to the model with the minimum AIC in the set ($\text{AIC}_{\min}$):
\cite{Wagenmakers2004, AKAIKE19781877, AKAIKE1979, Bozdogan_1987}:
\begin{align}
    \Delta_i(\text{AIC}) &= \text{AIC}_i - \text{AIC}_{\min} \label{eq:delta_aic}
\end{align}
The likelihood of model $\mathcal{M}_i$ being the best model (in the Kullback-Leibler information sense), given the data $\mathbf{x}$, can be estimated relative to the other models using these AIC differences. 
The relative likelihood of model $\mathcal{M}_i$, sometimes called an "Akaike weight", is given by \cite{Wagenmakers2004, burnham2002model}:
\begin{align}
    \mathcal{L}_{\rm rel}(\mathcal{M}_i|\mathbf{x}) &\propto \exp\left(-\frac{1}{2} \Delta_i(\text{AIC}) \right)
\end{align}
To obtain posterior probabilities for each model, $\mathcal{P}(\mathcal{M}_i|\mathbf{x})$, these relative likelihoods are normalized by summing over all models in the candidate set:
\begin{align}
    \mathcal{P}(\mathcal{M}_i|\mathbf{x}) = \frac{\exp\left(-\frac{1}{2} \Delta_i(\text{AIC})\right)}{\sum_{j=1}^{N} \exp\left(-\frac{1}{2} \Delta_j(\text{AIC})\right)} \label{eq:posterior_prob_delta_aic}
\end{align}
Substituting Eq. \eqref{eq:delta_aic} into Eq. \eqref{eq:posterior_prob_delta_aic}:
\begin{align}
\mathcal{P}(\mathcal{M}_i|\mathbf{x}) &= \frac{\exp\left(-\frac{1}{2} (\text{AIC}_i - \text{AIC}_{\min})\right)}{\sum_{j=1}^{N} \exp\left(-\frac{1}{2} (\text{AIC}_j - \text{AIC}_{\min})\right)} \\
&= \frac{\exp\left(-\frac{1}{2} \text{AIC}_i\right)}{\sum_{j=1}^{N} \exp\left(-\frac{1}{2} \text{AIC}_j\right)} \label{eq:posterior_prob_aic_i}
\end{align}
Now, substituting the definition of $\text{AIC}_i$ from Eq. \eqref{eq:aic_def} and under consideration that all models in \texttt{COMPASS} under comparison have the same number of parameters, i.e., $k_i = k_j = k$ for all $i,j$, then the additional parameter term is common to the numerator and all terms in the sum in the denominator, and thus cancels out. 
In this specific scenario, the posterior model probability simplifies to a direct ratio of the maximized likelihoods of the data given each model:
\begin{align}
    \mathcal{P}(\mathcal{M}_i|\mathbf{x}) &= \frac{\mathcal{L}(\mathbf{x}|\hat \theta_i, \mathcal{M}_i)}
    {\sum_{j=1}^{N} \mathcal{L}(\mathbf{x}|\hat \theta_j, \mathcal{M}_j )} \quad (\text{if } k_i = k_j \text{ for all } i,j)
\end{align}
The model with the highest posterior probability $\mathcal{P}(\mathcal{M}_i|\mathbf{x})$ is then considered the best-supported model by the data, according to this criterion.

\section{Network Architecture \& Calibration}
\label{sec:calibration}
\subsection{Conditional Transformer Architecture} \label{sub-sub-section:Results.Calibration.Network}

\texttt{COMPASS} uses a time-dependent transformer, \texttt{ConditionTransformer}, to approximate the conditional score function $s_\phi(\mathbf{z}_t, \mathcal{M}_C, t) \approx \nabla_{\mathbf{z}_t} \log p_t(\mathbf{z}_t)$, following advances in diffusion-based generative modeling \citep{peebles2023scalablediffusionmodelstransformers, gloeckler2024allinonesimulationbasedinference}.

Built on DiT \citep{peebles2023scalablediffusionmodelstransformers}, it applies adaLN-Zero and timestep embeddings, and takes as input $\mathbf{z} = (\theta, \mathbf{x}) \in \mathbb{R}^{D_\theta + D_x}$. A binary mask $\mathcal{M}_C \in \{0,1\}^{D}$ specifies which dimensions are observed ($1$) or latent ($0$).
Custom attention masks restrict each latent token to attend only to observed tokens, preventing leakage between latents. This supports two inference modes: 
\begin{itemize}
    \item \textbf{NPE}: Infer $\theta$ given $\mathbf{x}$ by masking $\mathcal{M}_C = (0, 1)$
    \item \textbf{NLE}: Infer $\mathbf{x}$ given $\theta$ by masking $\mathcal{M}_C = (1, 0)$
\end{itemize}

\subsection{Hyperparameter Tuning}
Key hyperparameters -- transformer depth, hidden size, MLP ratio, and attention heads --were optimized using \texttt{OPTUNA} \citep{akiba2019optunanextgenerationhyperparameteroptimization} across $1000$ trials to balance predictive accuracy $-\log P(\theta|x)$ and posterior coverage $\Delta_{\text{max}}$TARP. 

Our final architecture parameters are a batch size of 125, a sigma of 2.5, a depth of 5, 1 head and a hidden size of 65 with an MLP ratio of 3. 

\subsection{Diffusion Time} \label{sub-sub-section:Results.Calibration.Time}
\begin{figure}[!htb]
    \centering
    \includegraphics[width=.75\linewidth]{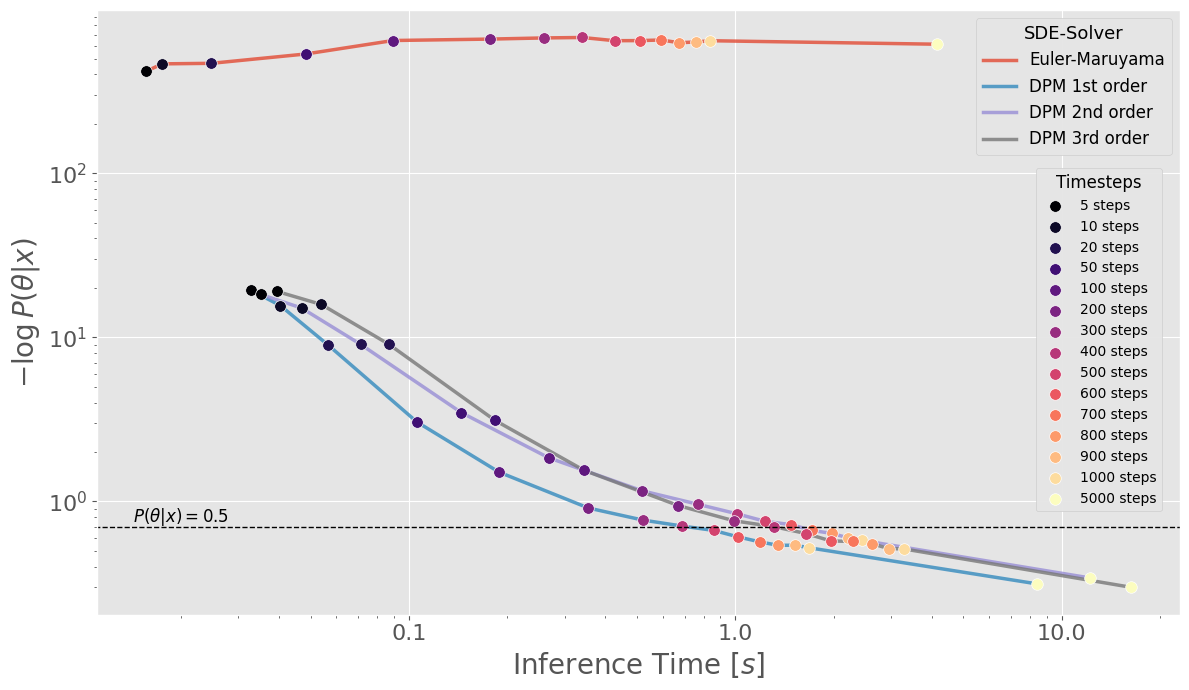}
\vspace{-1em}
\caption{Accuracy of SDE-Solvers 
Comparison of SDE solver performance in terms of predictive accuracy ($-\log P(\theta|x)$) versus inference time per sample. 
Euler-Maruyama and DPM-Solvers (1st, 2nd, and 3rd order) were tested with varying numbers of diffusion steps (indicated by colored points, ranging from 5 to 5000 steps).
Accuracy is averaged over $1000$ mock observations. The dashed horizontal line at $-\log P(\theta|x) \approx 0.693$ represents a posterior probability of $P(\theta|x) = 0.5$ for the true parameter. 
Lower $-\log P(\theta|x)$ values indicate higher accuracy. 
The DPM-Solver (1st order) with $500$ steps offers a good balance of accuracy and computational efficiency.}
\label{fig:AccuracyVsTime}
\end{figure}
Inference quality depends on reverse SDE discretization. We evaluated Euler-Maruyama and 1st–3rd order DPM-Solvers \citep{lu2022dpmsolverfastodesolver} over 15 timestep schedules, measuring $-\log P(\theta|x)$ vs. runtime on 1000 test samples (see Fig.~\ref{fig:AccuracyVsTime}).
The 1st-order DPM-Solver offers the best accuracy-efficiency trade-off, achieving reliable posteriors ($-\log P(\theta|x) < 0.693$) with 500 steps and 1s/sample inference time on 8× RTX 2080 Ti GPUs. This configuration is adopted as default.

\subsection{Sampling with COMPASS} \label{sec:COMPASS.Core.Sampling}
Once trained, $s_\phi$ is used to generate conditional samples from $p_0(\mathbf{z}_{\text{latent}}|\mathbf{z}_{\text{observed}})$ by solving the reverse SDE.
COMPASS employs DPM-Solver \citep{lu2022dpmsolverfastodesolver} to integrate the deterministic term of the VESDE:
\begin{align}
    \frac{d\mathbf{z}}{dt} = -\sigma^{2t} s_\phi(\mathbf{z}, \mathcal{M}_C, t)
\end{align}
The first-order update for step $t \to t'$ is:
\begin{align}
\mathbf{z}_{t'} = \mathbf{z}_t - (1 - \mathcal{M}_C) \, \sigma_t \, s_\phi(\mathbf{z}_t, \mathcal{M}_C, t) \, dt
\end{align}
Only latent dimensions are updated. Higher-order solvers improve this with intermediate evaluations.
To enhance fidelity, COMPASS adds optional Langevin corrector steps:
\begin{equation}
\mathbf{z} \leftarrow \mathbf{z} + \delta_L \, \sigma_{t'}^2 s_\phi(\mathbf{z}, \mathcal{M}_C, t') + \sqrt{2\delta_L \sigma_{t'}^2} \cdot \mathbf{n}, \quad \mathbf{n} \sim \mathcal{N}(0, \mathbf{I})
\end{equation}

With $\delta_L$ as the Langevin SNR, this stochastic refinement improves sample quality and avoids mode collapse. Correctors are applied every 5 steps (5 iterations per trigger, by default).
This predictor-corrector scheme combines the efficiency of deterministic solvers with stochastic refinement, enabling accurate, robust inference.

\subsection{Posterior Calibration}
\label{sub-sub-section:calibration} 
\begin{figure}[!htb]
\begin{minipage}{0.71\textwidth}
\centering
    \includegraphics[width=1\linewidth]{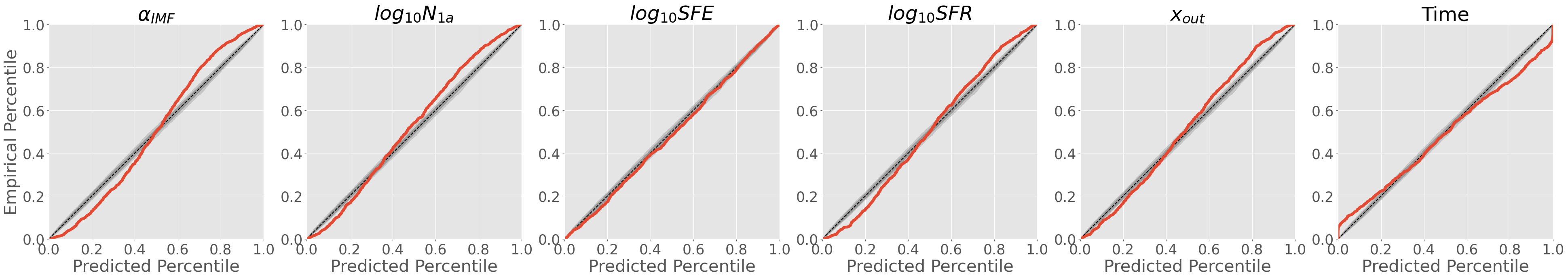}
    \includegraphics[width=1\linewidth]{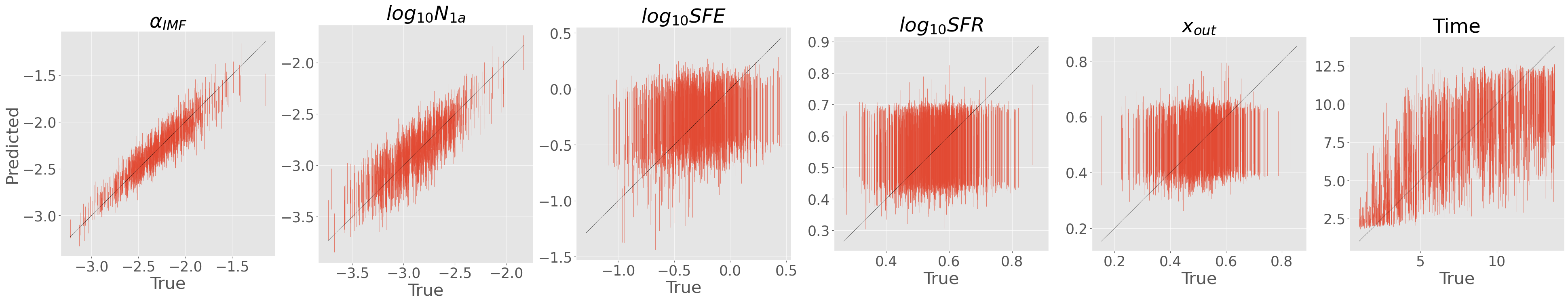}
\vspace{-2em}

\end{minipage}
\begin{minipage}{0.27\textwidth}
    \centering
    \includegraphics[width=\linewidth]{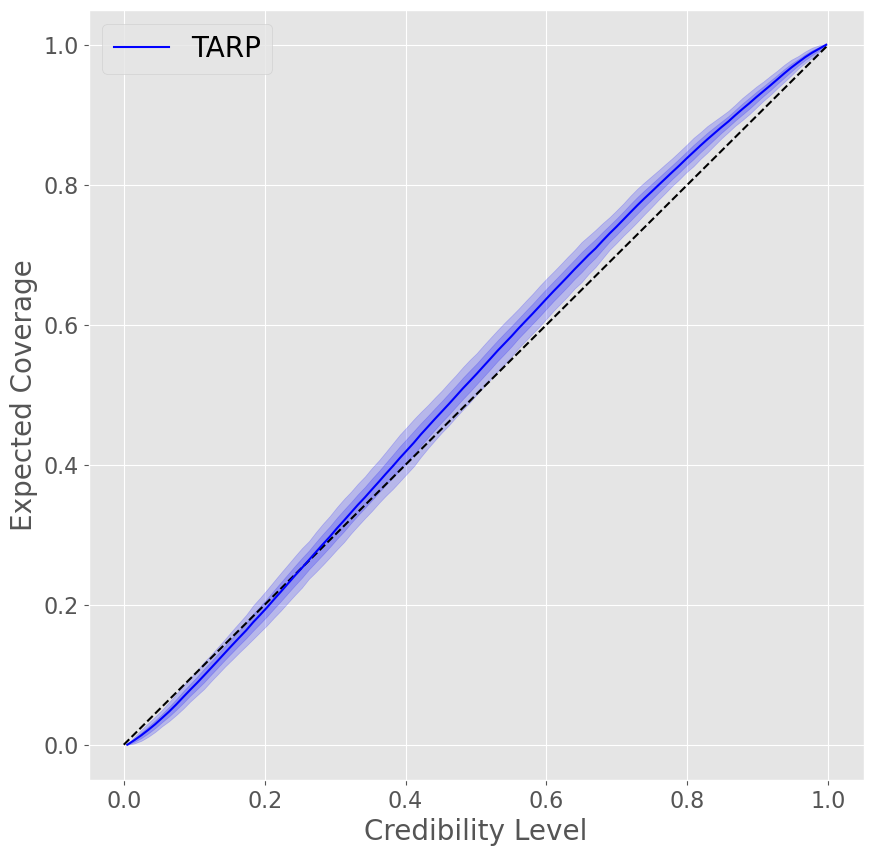}
\vspace{-1em}
\end{minipage}
\caption{Posterior calibration diagnostics showing the TARP plots at the top and the true vs. predicted parameter plots for each of the six parameters at the bottom and the TARP over all parameters on the right} \label{fig:Calibration}
\end{figure}
To ensure reliable uncertainty quantification, we evaluate posterior calibration using the TARP diagnostic and true-vs-predicted plots.
Figure~\ref{fig:Calibration} presents calibration results for the six \texttt{CHEMPY} parameters. TARP plots (top row) show credibility intervals match empirical coverage well—curves align with the diagonal, confirming well-calibrated uncertainty estimates. Posterior mean vs. true parameter plots (bottom row) show strong agreement for global parameters ($\alpha_{\rm IMF}$, $\log_{10}(N_{Ia})$), which benefit from strong data constraints.
Local parameters ($\log_{10}(\text{SFE})$, $\log_{10}(\text{SFR}_{\text{peak}})$, $x_{\rm out}$, and $T$) show greater spread, reflecting increased uncertainty and weaker constraints—expected given their spatial and temporal variability.

On the right of figure~\ref{fig:Calibration} is the aggregated TARP across all parameters and test samples. The curve closely tracks $y = x$, confirming that the overall posterior coverage is statistically sound. For example, 90\% intervals contain the true values approximately 90\% of the time.

In summary, the selected architecture and solver yield accurate, calibrated posteriors—especially for global parameters—supporting robust Bayesian inference and model comparison. The broader posteriors for local parameters reflect inherent data limitations.

\subsection{Significance test}
\label{sec:significance}
To assess whether the data supports a candidate model $\mathcal{M}_j$ over a simpler baseline $H_0$, we use a likelihood-ratio test comparing their marginal likelihoods:

\begin{align}
    K &= \frac{\mathcal{L}_{\mathcal{M}_j}(\mathbf{x})}{\mathcal{L}_{H_0}(\mathbf{x})} 
    \quad \text{or} \quad
    K = \log \mathcal{L}_{\mathcal{M}_j}(\mathbf{x}) - \log \mathcal{L}_{H_0}(\mathbf{x}) \label{eq:bayes-factor-log}
\end{align}

Here, $K$ is the \textit{Bayes Factor} \citep{MOREY20166}, quantifying how much more likely the observed data $\mathbf{x}$ is under model $\mathcal{M}_j$ than under $H_0$. 
$K > 1$ supports $\mathcal{M}_j$ while $K < 1$ supports $H_0$.
This Bayesian model comparison allows principled hypothesis testing beyond point estimates, reflecting both model fit and complexity \citep{Glover2004}.

\section{Testing on \texttt{CHEMPY} Mock Observational Data} \label{section:Results.Yields.Chempy}

To evaluate the model comparison capabilities of the \texttt{COMPASS} framework, an initial test was conducted using controlled mock observational data.
Three distinct combinations of nucleosynthetic yield sets were selected for this experiment, with the SN~Ia and AGB yields fixed to \citet{Seitenzahl2013} and \citet{Karakas2016}, respectively.
Only the core-collapse supernova (CC-SN) yields were varied among three widely used sets: \citet{Chieffi04} (Chieffi), \citet{Limongi_2018} (Limongi), and \citet{West17} (West), following the rationale in \citet{Buck2021}, who highlighted the significant variation in predicted yields across these models.

Bayesian model comparison involves estimating the model posterior probability $P(\mathcal{M}|x)$, which typically requires full posterior inference of the parameters $\theta$ under each candidate model $\mathcal{M}$, followed by evaluation of the likelihood $P(x|\theta, \mathcal{M})$.
This process can be computationally expensive if performed at high precision for every model.
However, when the primary goal is model selection rather than precise parameter estimation, a high-fidelity posterior $P(\theta|x, \mathcal{M})$ may not be necessary.
To reduce inference time while preserving sufficient accuracy for model discrimination, the number of diffusion timesteps was reduced.
Based on the results from Figure~\ref{fig:AccuracyVsTime}, using $50$ diffusion steps lowers the average inference time per sample to approximately $0.1$ seconds, while still maintaining an acceptable level of accuracy.

\begin{figure}[!htb]
    \centering
    \includegraphics[width=.6\linewidth]{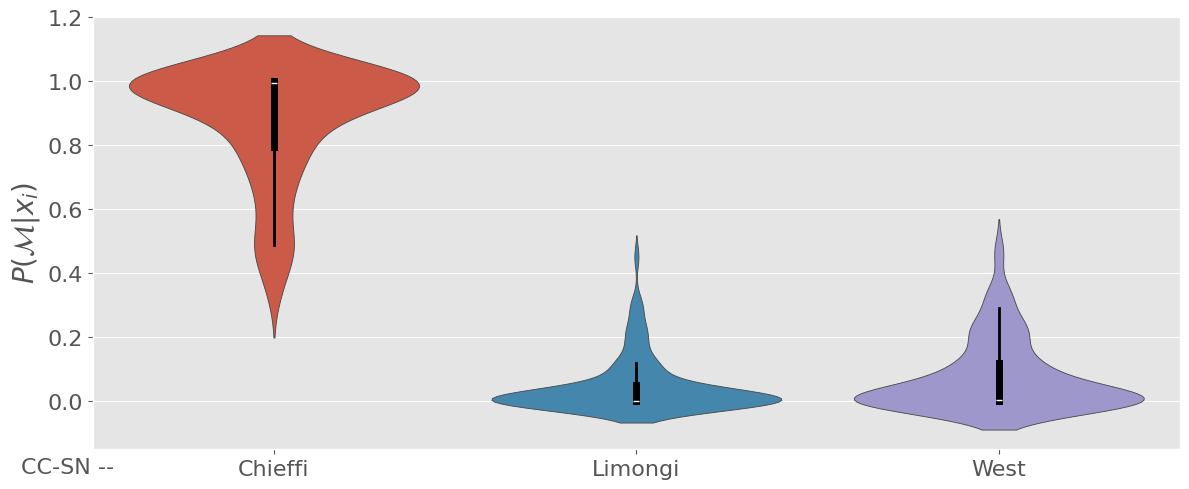}
    \includegraphics[width=.6\linewidth]{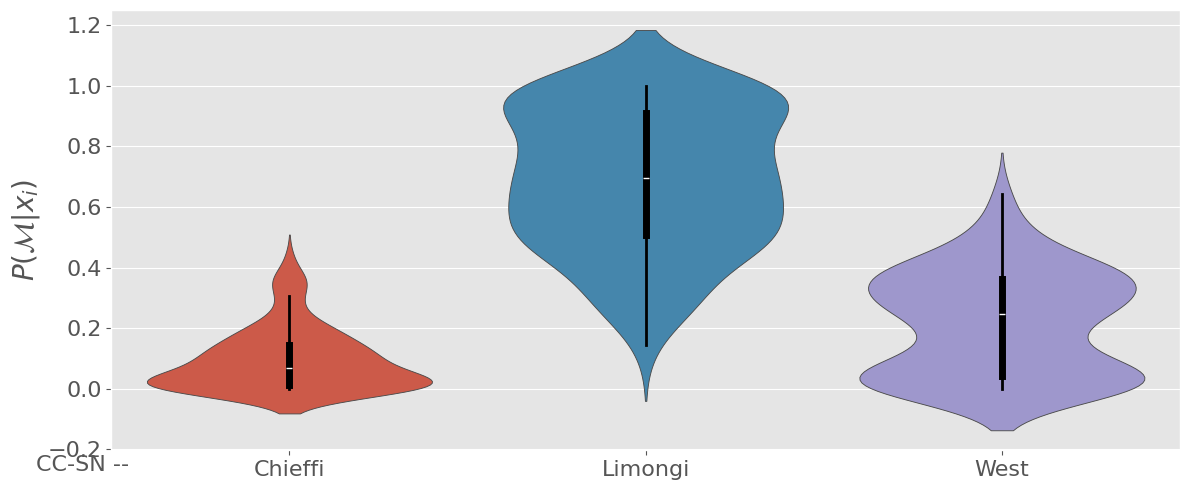}
    \includegraphics[width=.6\linewidth]{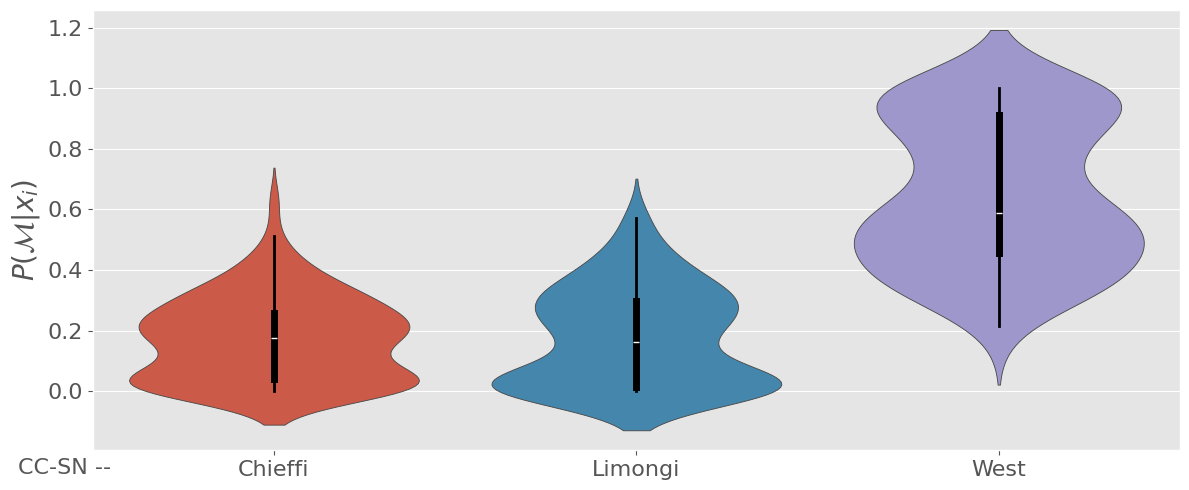}
\vspace{-1em}
\caption{Violin plots showing the distribution of single-observation posterior probabilities $P(\mathcal{M}_k|x_i)$ for three competing CC-SN yield models (Chieffi, Limongi and West), when tested on mock data.
Each panel corresponds to a different ground-truth model used to generate the mock observations:
\textbf{(Top):} Data generated with \citet{Chieffi04} yields.
\textbf{(Middle):} Data generated with \citet{Limongi_2018} yields.
\textbf{(Bottom):} Data generated with \citet{West17} yields.
The framework correctly identifies the true generating model with high probability in most individual observations, even with only $50$ diffusion timesteps.} \label{fig:MTf-tests}
\end{figure}

Figure~\ref{fig:MTf-tests} presents the resulting distributions of relative model posterior probabilities $P(\mathcal{M}_k|x_i)$ for individual mock observations $x_i$, where $\mathcal{M}_k$ represents one of the three CC-SN yield models.
Each of the three yield sets (Chieffi, Limongi, West) was used to generate a separate mock dataset, and the \texttt{COMPASS} framework was tasked with identifying the correct underlying model from the candidate set, using only $50$ diffusion timesteps.
The top panel of Figure~\ref{fig:MTf-tests} shows inference results for data generated using the Chieffi model; the middle panel corresponds to Limongi; and the bottom panel to West.
In each case, \texttt{COMPASS} correctly identifies the true generative model with high posterior probability for the majority of individual observations, even at reduced computational cost.

\begin{figure}[!htb]
    \centering
    \includegraphics[width=.75\linewidth]{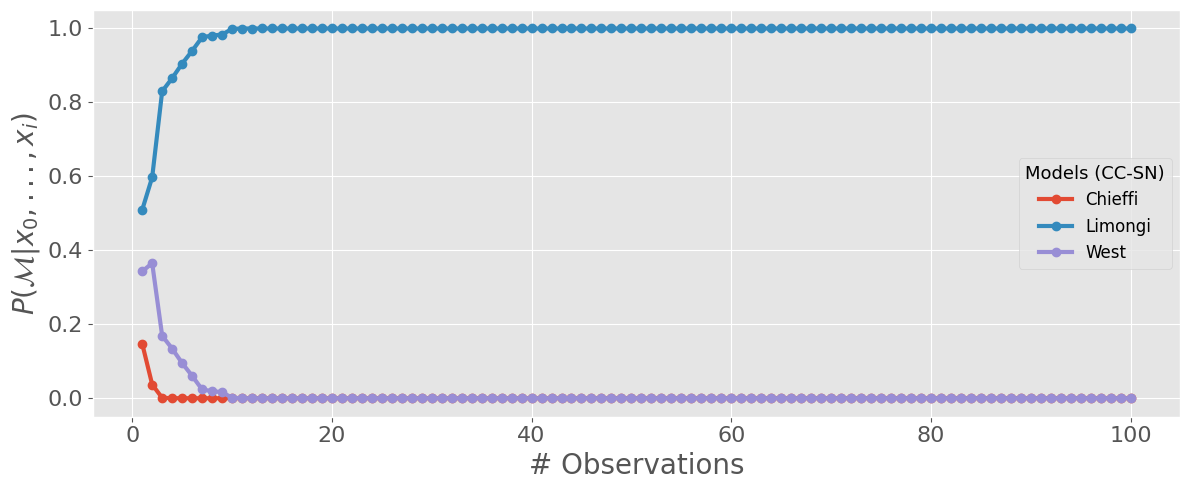}
\vspace{-1.5em}
\caption{Cumulative relative model posterior probability $P(\mathcal{M}_k|x_0,\dots,x_i)$ as a function of the number of combined mock observations $i$.
This example illustrates the case where the mock data was generated using the \citet{Limongi_2018} CC-SN yields (corresponding to the middle panel of Fig. \ref{fig:MTf-tests}). 
The probability rapidly converges to $100\%$ for the correct model (Limongi) after only $10$ observations, demonstrating the increased confidence gained by combining evidence from multiple data points.} \label{fig:MTf-test-updated}
\end{figure}

While individual posteriors provide strong evidence for model discrimination, combining results across multiple observations further increases the statistical confidence.
Figure~\ref{fig:MTf-test-updated} illustrates this cumulative effect, showing the aggregated posterior probability as a function of the number of observations, using the same test data from the middle panel (Limongi) in Figure~\ref{fig:MTf-tests}.
The cumulative curve demonstrates that, after just $10$ independent observations, the posterior probability assigned to the true model (Limongi) reaches $100\%$, indicating decisive support.

These results confirm that \texttt{COMPASS} can perform efficient and accurate Bayesian model selection, even with significantly reduced diffusion timesteps.
This makes the framework highly suitable for large-scale testing and comparative inference across astrophysical models without incurring prohibitive computational costs.

\section{Testing simulation-based Inference with \texttt{COMPASS}}
\label{sec:inference}
 
\begin{figure}[!htb]
    \centering
    \includegraphics[width=0.4\linewidth]{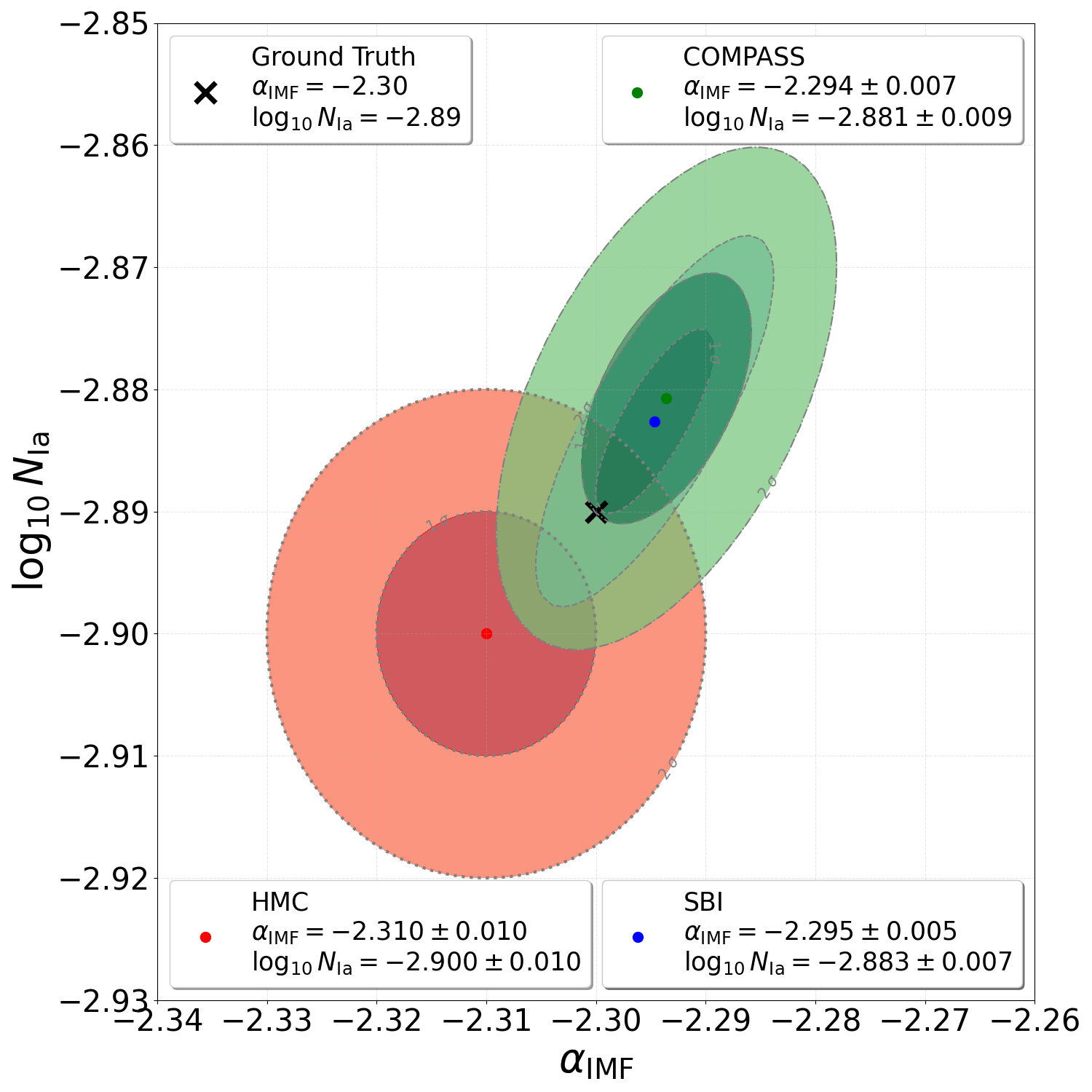}
    \includegraphics[width=0.4\linewidth]{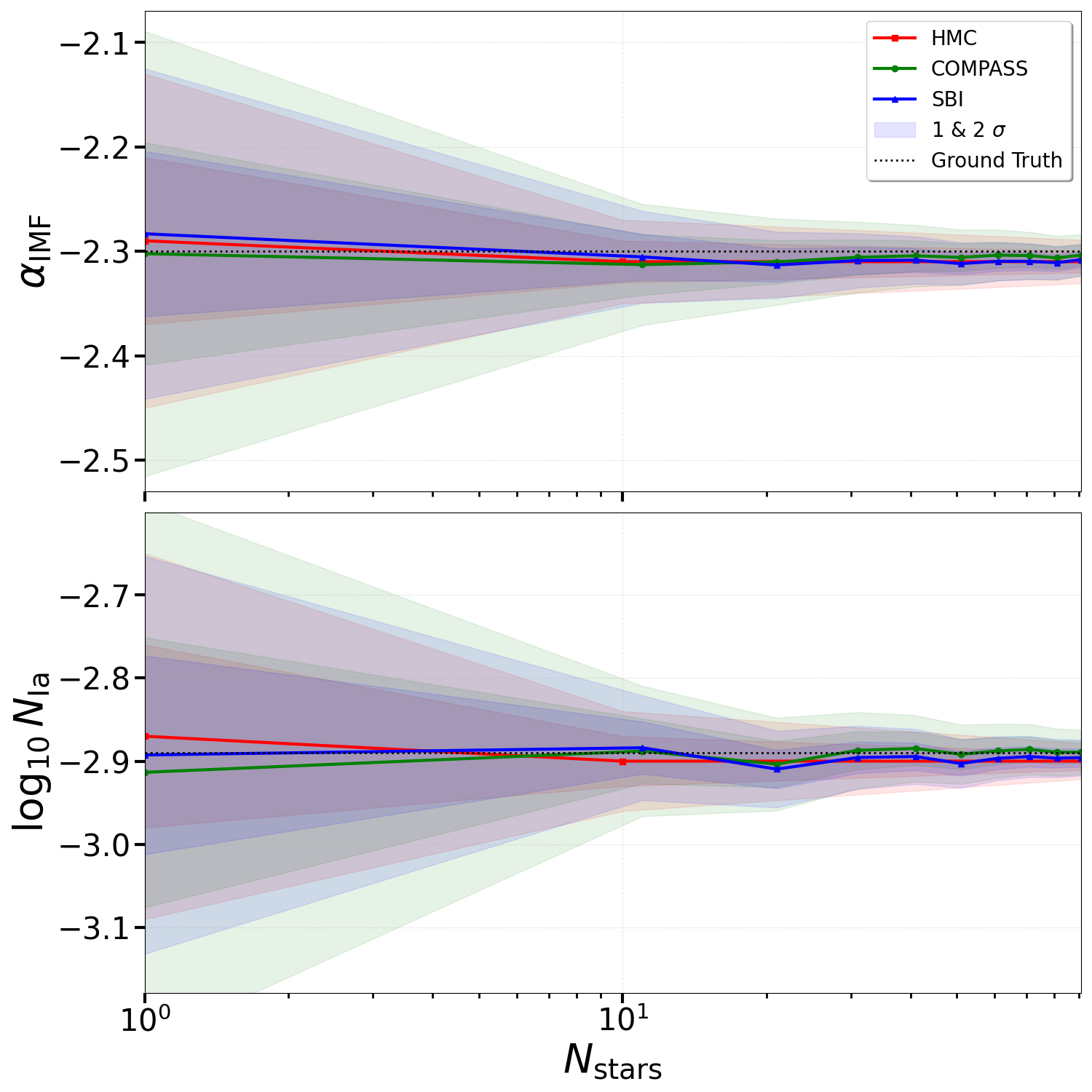}
\vspace{-1em}
\caption{Parameter Inference with Matched TNG Yields.
Comparison of inference results for the global galactic parameters $\Lambda$ using COMPASS (green), SBI (blue) and HMC (red) on mock data generated with TNG yields.
\textbf{(Left)}: Joint posterior distribution $P(\Lambda|\mathbf{x})$ inferred from $200$ stars.
\textbf{(Right):} Convergence of the inferred parameters as a function of $N_{\rm stars}$.} \label{fig:sbi-chempy-TNG}

    \centering
    \includegraphics[width=0.4\linewidth]{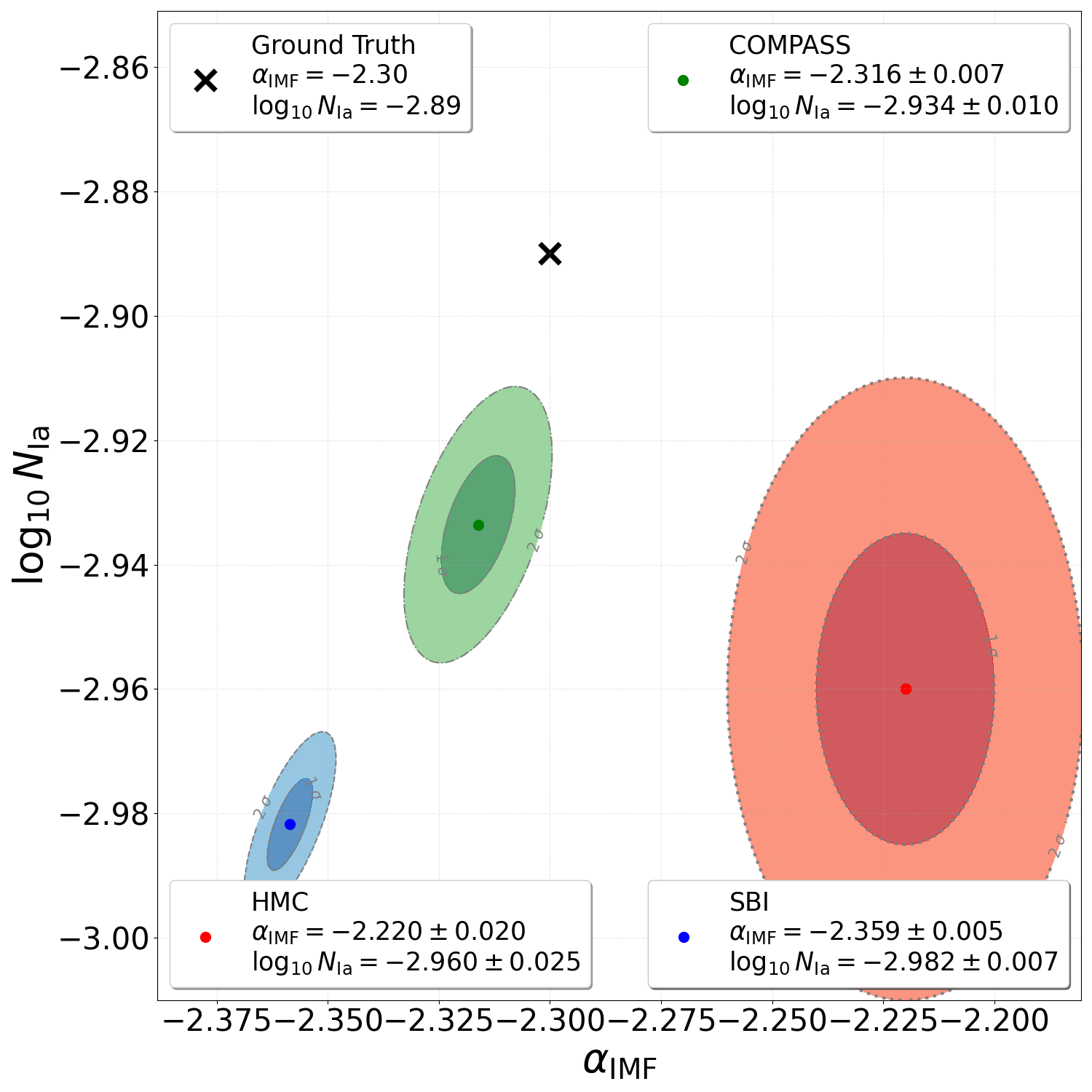}
    \includegraphics[width=0.4\linewidth]{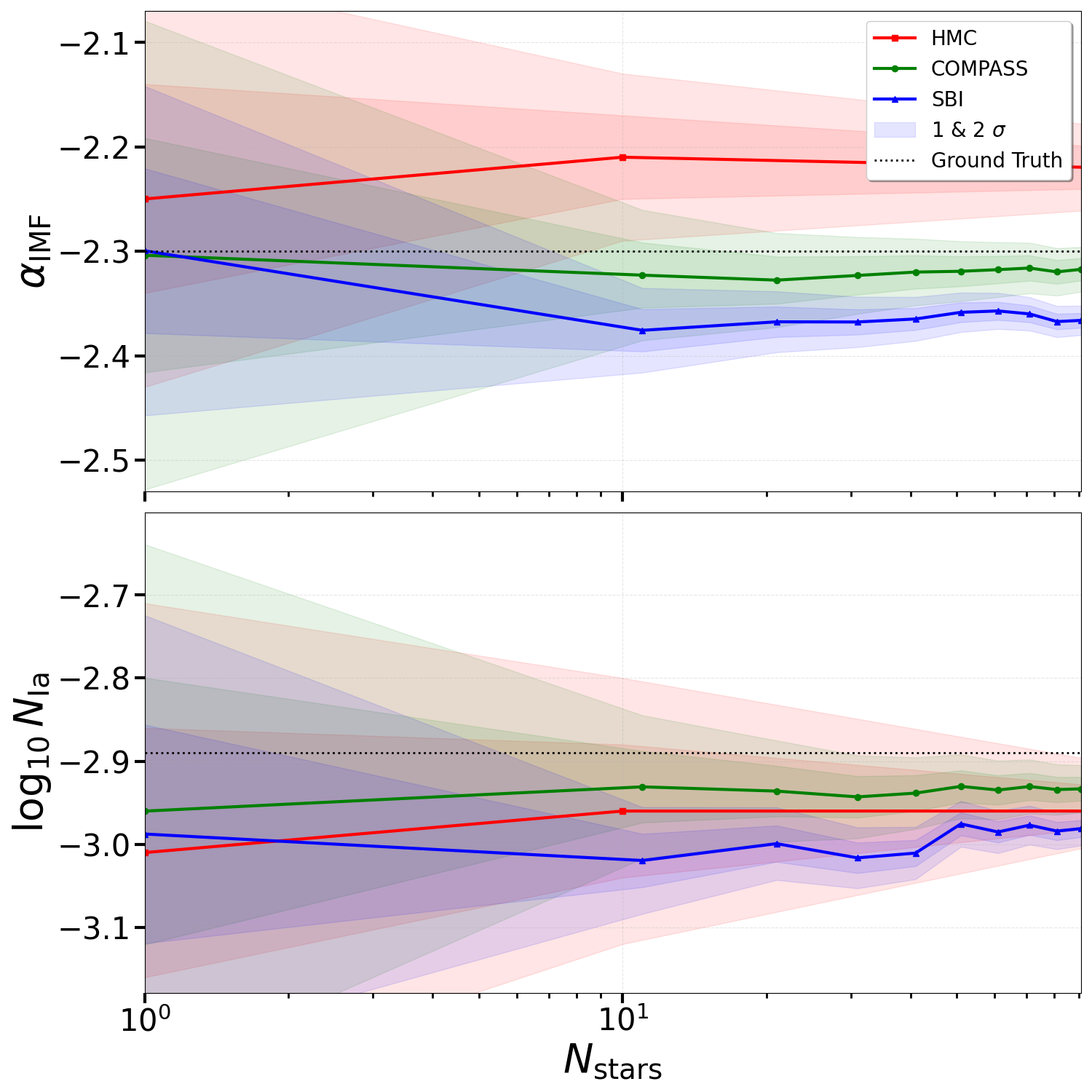}
\vspace{-1em}
\caption{Parameter Inference Alternative Yields
Comparison of inference results for the global galactic parameters $\Lambda$ using COMPASS (green), SBI (blue) and HMC (red) on mock data generated with the alternative yield sets (Tab. \ref{table:chempy_ALT_yields})
\textbf{(Left):} Joint posterior distribution $P(\Lambda|\mathbf{x})$ inferred from $200$ stars. 
\textbf{(Right):} Convergence of the inferred parameters as a function of $N_{\rm stars}$.} \label{fig:sbi-chempy-alt}

    \centering
    \includegraphics[width=0.4\linewidth]{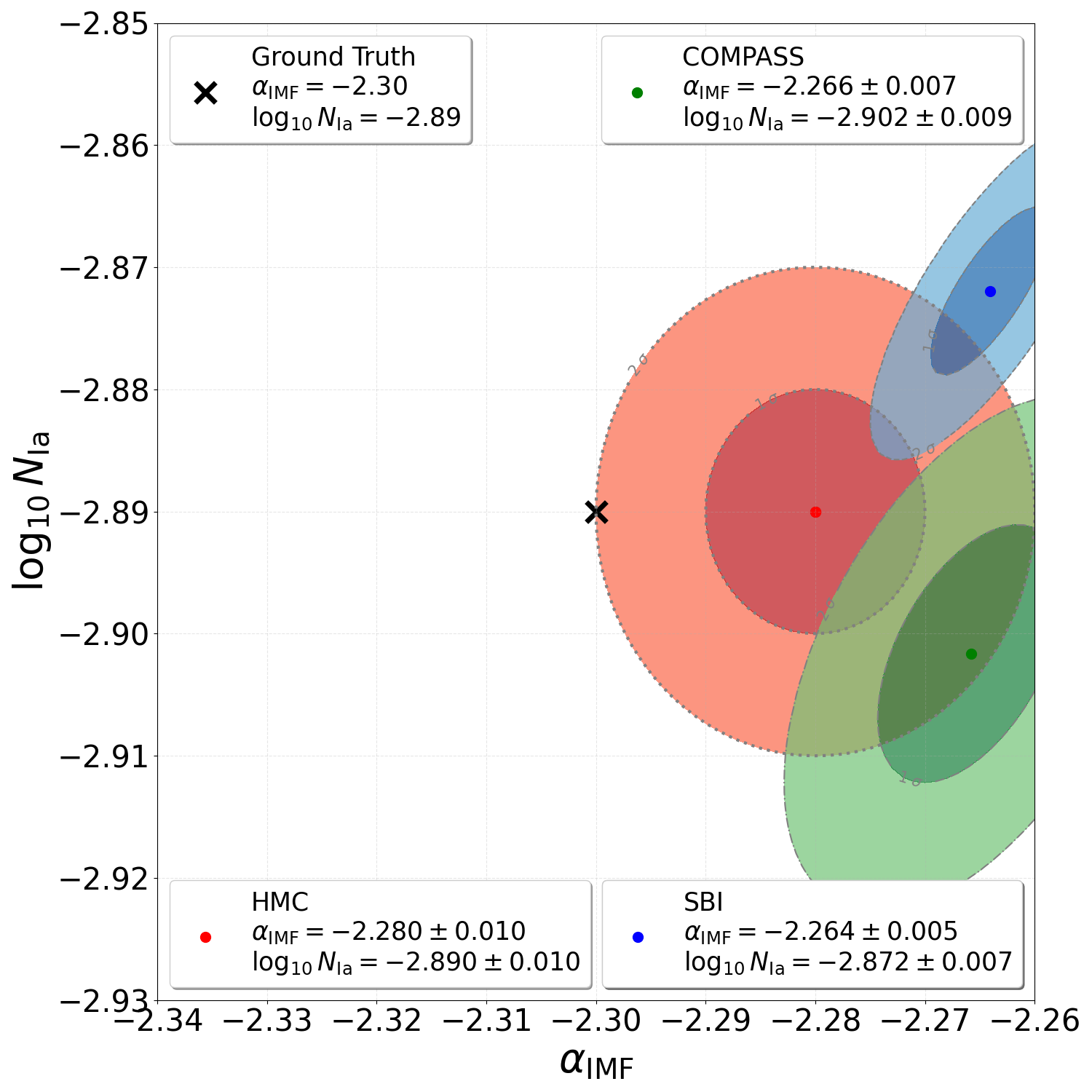}
    \includegraphics[width=0.4\linewidth]{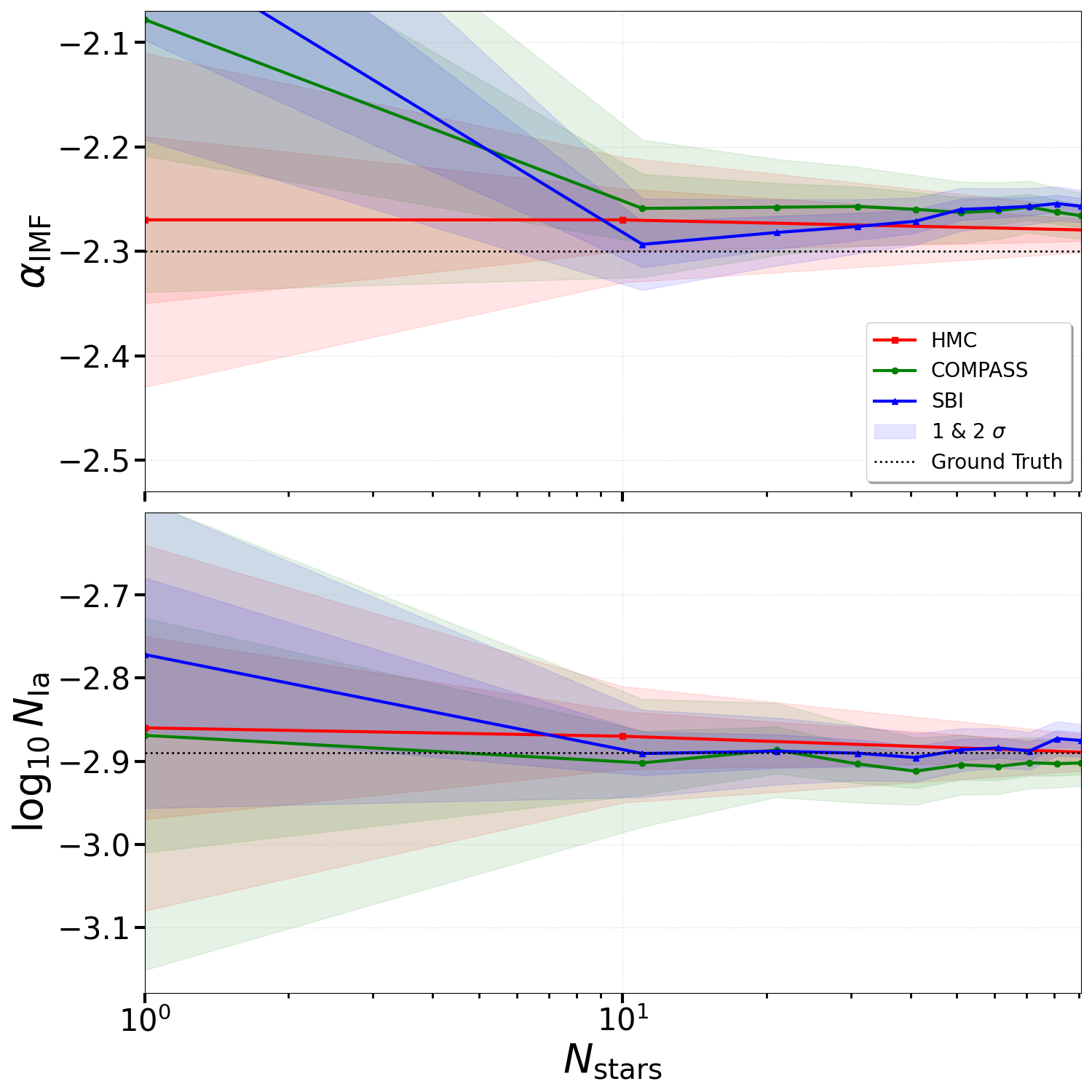}
\vspace{-1em}
\caption{Parameter Inference from IllustrisTNG Simulation
Comparison of inference results for the global galactic parameters $\Lambda$ using COMPASS (green), SBI (blue) and HMC (red) on stellar abundances from the IllustrisTNG simulated galaxy.
\textbf{(Left):} Joint posterior distribution $P(\Lambda|\mathbf{x})$ inferred from $200$ stars. 
\textbf{(Right):} Convergence of the inferred parameters as a function of $N_{\rm stars}$.} \label{fig:sbi-chempy-TNG-sim}

\end{figure}

To rigorously assess the parameter inference capabilities of the \texttt{COMPASS} framework using Simulation-Based Inference (SBI), a series of tests were conducted with mock observational data generated by the one-zone Galactic Chemical Evolution (GCE) simulator \texttt{CHEMPY}. 
The primary objective is to benchmark \texttt{COMPASS} against the SBI methodology from \citet{buck2025inferringgalacticparameterschemical}, which employs a Neural Posterior Estimator (NPE), and also against traditional Hamiltonian Monte Carlo (HMC) inference pipeline developed by \citep{Philcox_2019}, which serves as the direct methodological predecessor to this work. 
This allows for evaluating whether \texttt{COMPASS} can reliably recover known ground-truth parameters under various inference scenarios.

All inference models discussed in this section were trained on the IllustrisTNG nucleosynthetic yield tables implemented in \texttt{CHEMPY} (see Table~\ref{table:chempy_TNG_yields}).
\paragraph{TNG Yield Sets}
For initial validation, mock data were generated using the same TNG yield tables as used in the training phase (see Table~\ref{table:chempy_TNG_yields}). 
The global parameters were fixed at $\alpha_{\mathrm{IMF}} = -2.3$ and $\log_{10}N_{\mathrm{Ia}} = -2.89$.
Figure~\ref{fig:sbi-chempy-TNG} compares inference results from \texttt{COMPASS}, the SBI pipeline from \citet{buck2025inferringgalacticparameterschemical}, and the HMC method of \citet{Philcox_2019}.

As seen in Figure~\ref{fig:sbi-chempy-TNG}, both \texttt{COMPASS} and SBI accurately recover the ground-truth parameters, with precision improving as more stars are included in the sample. 
Thanks to their amortized inference nature, both SBI and \texttt{COMPASS} can scale to larger sample sizes with minimal computational overhead -- an important feature given the high sample variance when using small stellar datasets.
\texttt{COMPASS} demonstrates inference accuracy comparable to the previous SBI implementation, with deviations of less than $0.5\%$ for both $\alpha_{\mathrm{IMF}}$ and $\log_{10}N_{\mathrm{Ia}}$, and the ground truth consistently lying within the $1\sigma$ credible interval.
This highlights \texttt{COMPASS}'s utility not only for model comparison, but also for precise parameter inference.

\begin{table}[H]
    \begin{minipage}{0.5\textwidth}
        \centering
        \begin{tabular}{c|c}
            \toprule
            \textbf{Type} & \textbf{Yield Table} \\
            \midrule
            SN Ia & \citet{1997NuPhA.621..467N}\\
            CC-SN & \citet{Kobayashi_2006}; \citet{portinari1997galacticchemicalenrichmentnew}\\
            AGB & \thead{\citet{2010MNRAS.403.1413K}; \citet{2014MNRAS.437..195D}\\ \citet{2014ApJ...797...44F}} \\
            \bottomrule
        \end{tabular}
    \caption{TNG Yield Tables} \label{table:chempy_TNG_yields}
    \end{minipage}
    \begin{minipage}{0.5\textwidth}
         \centering
         \begin{tabular}{c|c}
            \toprule
            \textbf{Type} & \textbf{Yield Table} \\
            \midrule
            SN Ia & \citet{2003NuPhA.718..139T} \\
            CC-SN & \citet{Nomoto2013} \\
            AGB & \citet{Karakas2016}\vphantom{\thead{\citet{2010MNRAS.403.1413K}; \citet{2014MNRAS.437..195D}\\ \citet{2014ApJ...797...44F}}} \\
            \bottomrule
         \end{tabular}
     \caption{Alternative Yield Tables} \label{table:chempy_ALT_yields}
     \end{minipage}
\end{table}

\paragraph{Alternative Yield-Sets}
Since no tabulated nucleosynthetic yield set perfectly reflects nature, and it is often unclear which yield combination is most realistic, this section investigates the sensitivity of inference results to yield set misspecification.
To that end, mock data were generated using an alternative yield set (Table~\ref{table:chempy_ALT_yields}) while still fixing the global parameters at $\alpha_{\mathrm{IMF}} = -2.3$ and $\log_{10}N_{\mathrm{Ia}} = -2.89$.
The local ISM parameters $(\Theta_i, T)$ were drawn from their priors.
These alternative yields were chosen such that each nucleosynthetic channel differs by approximately $\mathcal{O}(10\%)$ in predicted abundance yields, providing a meaningful test of model misspecification.

Figure~\ref{fig:sbi-chempy-alt} presents inference results under these alternative yields.
All three inference methods show degradation in accuracy due to the incorrect model assumptions, but \texttt{COMPASS} maintains closer alignment with the ground-truth values.
These results reinforce the importance of correct yield-set selection for reliable parameter inference.

\paragraph{IllustrisTNG Simulation}
While \texttt{CHEMPY} provides a computationally efficient one-zone GCE framework, it simplifies several key physical processes in the interstellar medium (ISM), such as gas mixing, starbursts, and environmental coupling.
To evaluate whether these simplifications bias parameter inference, a more physically realistic dataset was constructed from a full hydrodynamical simulation.
Specifically, a Milky Way-like galaxy was selected from the $z=0$ snapshot of the high-resolution TNG100-1 simulation. 
Subhalo index $5223071$—with a halo mass close to $10^{12}\,M_\odot$—was chosen to represent a Milky Way analog.
From this system, $1{,}000$ stellar particles were randomly selected from a total of $\sim40{,}000$ available. 
Each stellar particle represents a population of stars and carries mass-weighted elemental mass fractions $\{d_i^j\}$ and a formation time given by the cosmological scale factor $a_i$.

The elemental abundances were converted to [X/Fe] using the solar reference values from \citet{2009ARA&A..47..481A}, consistent with \texttt{CHEMPY}. 
Formation times $T_i$ were derived from $a_i$ using the \texttt{astropy} cosmology package \citep{astropy:2013,astropy:2018}, assuming a flat $\Lambda$CDM model with parameters from \citet{planck2015}, as adopted in the TNG simulations.
To ensure compatibility with the neural network training regime, particles with $T_i \notin [2, 12.8]$ Gyr were excluded, removing roughly $5\%$ of the dataset.
The final dataset mirrors the structure of the \texttt{CHEMPY} mock data, with observational uncertainties included in the same way.

This particular TNG galaxy was chosen by \citet{Philcox_2019} to contain a clear bimodal distribution in $\alpha$-abundances—featuring both high- and low-$\alpha$ sequences—analogous to the Milky Way.
The origin of this chemical bimodality is still debated and has been linked to various formation scenarios, including gas-rich mergers, starbursts \citep[e.g.][]{2018MNRAS.474.3629G,2018MNRAS.477.5072M,2019MNRAS.484.3476C,Buck2020,Buck2023}, and radial migration with selection effects \citep[e.g.][]{2009MNRAS.396..203S,2013A&A...558A...9M,2017ApJ...835..224A}.

Inference results using this TNG galaxy data are shown in Figure~\ref{fig:sbi-chempy-TNG-sim}.
The results confirm that increasing the number of stellar particles improves inference accuracy.
Despite the mismatch between the complex TNG chemical enrichment model and the simpler training model (\texttt{CHEMPY}), both SBI and \texttt{COMPASS} successfully recover the underlying parameters.
The posterior for $\log_{10}(N_{\mathrm{Ia}})$ is nearly unbiased, while $\alpha_{\mathrm{IMF}}$ shows a slight overestimation across all methods.
Nonetheless, \texttt{COMPASS} and SBI deliver performance on par with HMC, and do so with significantly reduced computational cost, underscoring their practical advantage.

In summary, Figures~\ref{fig:sbi-chempy-TNG}, \ref{fig:sbi-chempy-alt}, and \ref{fig:sbi-chempy-TNG-sim} collectively demonstrate the robustness of simulation-based inference for galactic parameter estimation from stellar chemical abundances. When interpreting these results, it is crucial to consider the methodological assumptions underlying the uncertainty quantification. Both COMPASS and the benchmark SBI pipeline rely on a Gaussian approximation of individual stellar posteriors to facilitate their analytical combination. This posterior factorization, while computationally efficient, can lead to underestimated uncertainties and over-confident constraints in the regime of large stellar samples by not fully capturing the tails of the true posterior distributions.

We find that COMPASS consistently produces broader and thus more conservative posterior contours than the benchmark SBI pipeline. This is a direct consequence of our framework's explicit marginalization over observational uncertainties, a step that more realistically propagates measurement error into the final parameter constraints. While the Gaussian approximation remains a shared limitation, the overall accuracy of parameter recovery by COMPASS, combined with its more robust uncertainty estimation, provides strong validation for its use in both model selection and parameter inference.

\section{Additional information in inference results}

\definecolor{Nugrid-TNG}{HTML}{BE604F}
\definecolor{Nugrid-Chieffi}{HTML}{5584A8}
\definecolor{Karakas-Chieffi}{HTML}{9D97C8}
\definecolor{Ventura-TNG}{HTML}{777777}
\definecolor{Nugrid-ChieffiNet}{HTML}{DFBE7C}
\definecolor{Nugrid-Nomoto}{HTML}{91AA5D}
\begin{table}[!htb]
    \centering
    \resizebox{\columnwidth}{!}{%
    \begin{tabular}{l >{\columncolor{white}}c >{\columncolor{white}}c >{\columncolor{white}}c >{\columncolor{white}}c >{\columncolor{white}}c >{\columncolor{white}}c >{\columncolor{white}}c >{\columncolor{white}}c >{\columncolor{white}}c >{\columncolor{white}}c}
        \toprule
        \textbf{AGB Yields} & \multicolumn{10}{c}{\textbf{Core-Collapse Supernovae (CC-SN) Yields}} \\
        \cmidrule(lr){2-11} 
         & Chieffi & Nomoto  & Portinari & Chieffi Net & Nomoto Net & NuGrid  & West & TNG& CL18& Frischknecht \\
        \midrule
        NuGrid  & \cellcolor{Nugrid-Chieffi!70} \textbf{1} & 6  & 29 & \cellcolor{Nugrid-ChieffiNet!70} 4  & \cellcolor{Nugrid-Nomoto!70} 5  & 22 & 19 & \cellcolor{Nugrid-TNG!70} \textbf{0} & 30 & 36 \\
        Karakas & \cellcolor{Karakas-Chieffi!70} \textbf{2} & 18 & 31 & 14 & 21 & 26 & 23 & 9  & 35 & 39 \\
        Ventura & 10 & 16 & 34 & 13 & 17 & 27 & 20 & \cellcolor{Ventura-TNG!70} 3  & 32 & 38 \\
        TNG     & 7  & 15 & 28 & 11 & 12 & 25 & 24 & 8  & 33 & 37 \\
        \bottomrule
    \end{tabular}
    }
\caption{Ranking of Nucleosynthetic Yield Set Combinations Based on Bayesian Model Comparison with \citet{Nissen_2020} Data. Lower ranks indicate higher posterior probability and correspond to the numbers and colours in Figure \ref{fig:MTf-nissen}.} \label{table:yield-comparison-results}
\end{table}

\begin{table}[!htb]
    \centering
    \resizebox{\columnwidth}{!}{%
    \begin{tabular}{lcccccc}
        \toprule
        \thead{\textbf{AGB} \\ \textbf{CC-SN}} & \thead{Nugrid \\ TNG} & \thead{Nugrid\\ Chieffi} & \thead{Karakas\\Chieffi} & \thead{Ventura\\TNG} & \thead{Nugrid\\ Chieffi Net} & \thead{Nugrid\\Nomoto} \\
        \midrule
        $\alpha_{\rm IMF}$ & $-2.60 \pm 0.02$ & $-2.52 \pm 0.02$ & $-2.57 \pm 0.01$ & $-2.71 \pm 0.01$ & $-2.53 \pm 0.02$ & $-2.70 \pm 0.01$ \\
        $\log_{10}N_{\rm Ia}$ & $-2.78 \pm 0.02$ & $-2.70 \pm 0.02$ & $-2.83 \pm 0.02$ & $-3.00 \pm 0.02$ & $-2.71 \pm 0.02$ & $-2.86 \pm 0.02$ \\
        \bottomrule
    \end{tabular}}
\caption{Inferred Galactic Parameters from Observational Data.
Mean values and $\pm1\sigma$ uncertainties for $\alpha_{\rm IMF}$ and $\log_{10}(N_{\rm Ia})$ for the top six yield set combinations.} \label{table:nissen-inference-results}
\end{table}
%


\end{document}